\documentstyle[twocolumn,eqsecnum,aps]{revtex}
\begin{document}
\draft
\title{Trivial Vacua, High Orders in Perturbation Theory and Nontrivial
Condensates}
\author{M. Burkardt}
\address{
Physics Dept.\\
New Mexico State University\\
Las Cruces, NM 88003-0001}
\maketitle
\begin{abstract}
In the limit of an infinite number of colors,
an analytic expression for the quark condensate
in $QCD_{1+1}$ is derived as a function of the quark mass and the
gauge coupling constant. For zero quark mass, a nonvanishing
quark condensate is obtained. Nevertheless, it is shown
that there is no phase transition as a function of the
quark mass. It is furthermore shown that the expansion
of $\langle 0 | \overline{\psi}\psi |0\rangle$ in the gauge
coupling
has zero radius of convergence but that the perturbation
series is Borel summable with finite radius of convergence.
The nonanalytic behavior
$\langle 0 | \overline{\psi}\psi |0\rangle
\stackrel{m_q\rightarrow0}{\sim} - N_C \sqrt{G^2}$ can only be
obtained by summing the perturbation series to infinite order.
The sum-rule calculation is based on masses and coupling
constants calculated from 't Hooft's solution to
$QCD_{1+1}$ which employs LF quantization and is thus based
on a trivial vacuum. Nevertheless the chiral condensate
remains nonvanishing in the chiral limit which is yet another example that
seemingly trivial LF vacua are {\it not} in conflict with QCD sum-rule results.

\end{abstract}

\narrowtext
\section{Introduction}
What is interesting about the quark condensate
in $QCD_{1+1}$? Zhitnitsky \cite{zhit:vac}, using QCD-sum rule
techniques, derived an exact result for the
condensate in the limit of an infinite number
of colors\footnote{Note that Coleman's theorem \cite{co:no}
prohibits spontaneous breakdown of chiral
symmetry for any finite number of colors, since this
is a $1+1$-dimensional model.}
and in the limit $m_q\rightarrow 0$
\begin{equation}
\left.\langle 0 | \overline{\psi}\psi |0\rangle
\right|_{m_q=0} =
- \frac{N_C}{\sqrt{12}}\sqrt{\frac{g^2C_F}{\pi}},
\label{eq:mis0}
\end{equation}
where $C_F=(N_C^2-1)/2N_C$ and $G^2 \equiv \frac{g^2C_F}{\pi}$
is held fixed as $N_C\rightarrow \infty$.
This result is remarkable in several respects:
Firstly, $\left. \langle 0 | \overline{\psi}\psi |0\rangle
\right|_{m_q=0}$ is
nonvanishing, indicating spontaneous breakdown
of chiral symmetry. Secondly, the condensate is nonanalytic
in the coupling constant $G^2$, thus indicating
nonperturbative effects: although it seems natural from
dimensional analysis that $\langle 0 | \overline{\psi}\psi |0\rangle \propto
\sqrt{G^2}$ for small quark masses, it
is impossible to obtain such a behavior
in perturbation theory where one can only
generate terms $\propto G^{2n}$. Thirdly, one may suspect
that there is a phase transition in $QCD_{1+1}$.

Since the coupling constant $g$ in $QCD_{1+1}$ carries the dimension
of a mass, the theory is super-renormalizable and the
scale is set both by the coupling and by the mass.
In practice this implies that
$\langle 0 | \overline{\psi}\psi |0\rangle$
can (up to some dimensionful overall factor)
depend on $G^2$ only through the combination
$\alpha \equiv G^2/m_q^2$.
Therefore, in order to address the abovementioned issues
of nonperturbative effects and a possible phase
transition, it is necessary to consider nonzero quark masses.

There is another reason why the condensate in $QCD_{1+1} (N_C\rightarrow
\infty)$
is interesting: Zhitnitsky's result was based on the solutions
of 't Hooft's equation \cite{thooft}, which is obtained in the
light-front (LF) approach to $QCD_{1+1}$. As is well known,
the vacuum (=ground state) in LF quantization is equal to
the Fock vacuum and nontrivial condensates seem impossible.
However, as has been shown in Ref.\cite{zhit:vac}, QCD sum rules,
applied to the spectrum and coupling constants obtained through
LF quantization, nevertheless yield nontrivial results for
the condensates.

\section{$\langle 0|\bar{\psi}\psi|0\rangle$ from sum rules}\footnote{Fragments
of this calculation can be found in Ref.\cite{mb:tbp}.}
For nonzero quark masses, the vacuum expectation value
of the scalar density diverges already for free fields
\begin{equation}
\langle 0|\bar{\psi}\psi |0\rangle = \frac{N_C}{2\pi}
m_q\ln \frac{\Lambda^2}{m_q^2}
{}.
\label{eq:freediv}
\end{equation}
However, due to the mild UV behavior in $1+1$ dimensions
($QCD_{1+1}$ is super-renormalizable) it is sufficient to
subtract the free field expectation value (i.e. to
``normal order'') to render
$\langle 0|\bar{\psi}\psi |0\rangle$ finite. This motivates
the definition
\begin{equation}
\left. \langle 0|\bar{\psi}\psi |0\rangle \right|_{ren}
\equiv
\langle 0|\bar{\psi}\psi |0\rangle -
\left. \langle 0|\bar{\psi}\psi |0\rangle \right|_{g=0}.
\label{eq:subtract}
\end{equation}
The condensate itself can be evaluated using current algebra
\begin{eqnarray}
0 &=& \lim_{q \rightarrow 0} iq^\mu \int d^2x e^{iqx}
\langle 0|T\left[\bar{\psi}\gamma_\mu \gamma_5\psi (x)
\bar{\psi}i\gamma_5\psi (0)\right] |0\rangle
\nonumber\\
&=&-\langle 0|\bar{\psi}\psi |0\rangle
-2m_q \int d^2x
\langle 0|T\left[\bar{\psi}i \gamma_5\psi (x)
\bar{\psi}i\gamma_5\psi (0)\right] |0\rangle .
\nonumber\\
\label{eq:cural}
\end{eqnarray}
Upon inserting a complete set of meson states\footnote{
Because we are working at leading order in $1/N_C$, the
sum over one meson states saturates the operator product
in Eq.(\ref{eq:cural}).}
one thus obtains
\begin{equation}
\langle 0|\bar{\psi}\psi |0\rangle
= -m_q \sum_n \frac{f_P^2(n)}{M_n^2},
\label{eq:naivesum}
\end{equation}
where
\begin{equation}
f_P(n)\equiv \langle 0|\bar{\psi}i\gamma_5\psi |n\rangle
=\sqrt{\frac{N_C}{\pi}} \frac{m_q}{2}\int_0^1 dx \frac{1}{x(1-x)} \phi_n(x)\ \
\end{equation}
and the wavefunctions $\phi_n$ and invariant masses $M_n^2$ are
obtained from solving
`t Hooft's bound state equation for mesons in
$QCD_{1+1}$
\begin{equation}
M_n^2 \phi_n(x) = \frac{m_q^2}{x(1-x)}\phi_n(x) + G^2 \int_0^1 dy
\frac{\phi_n(x)-\phi_n(y)}{(x-y)^2}.
\label{eq:thooft}
\end{equation}
The variable $x$ corresponds to the light-front
momentum fraction carried by the quark in the meson.
Note that `t Hooft's equation was been derived using
light-front quantization --- we will return to this point below.

In the limit of highly excited mesons, the masses and coupling
constants scale \cite{ei:qcd}:
\begin{eqnarray}
M_n^2 &\stackrel{n\rightarrow \infty}{\longrightarrow}n\pi^2 G^2\nonumber\\
f_P(n) &\stackrel{n\rightarrow \infty}{\longrightarrow}\sqrt{N_C \pi G^2}
\end{eqnarray}
and thus the sum in Eq.(\ref{eq:naivesum}) diverges
logarithmically. Of course this only reflects the free field divergence
(\ref{eq:freediv}). In order to regularize Eq.(\ref{eq:naivesum})
in a gauge invariant way we introduce an invariant mass cutoff
and obtain
\begin{equation}
\left.\langle 0|\bar{\psi}\psi |0\rangle \right|_{ren}
= -m_q \lim_{\Lambda \rightarrow \infty}
\left[\sum_n \frac{f_P^2(n)}{M_n^2\left(1+M_n^2/\Lambda^2\right)}
-``g=0''\right].
\label{eq:regsum}
\end{equation}
Eq.(\ref{eq:regsum}) can be used to calculate
$\left.\langle 0|\bar{\psi}\psi |0\rangle \right|_{ren}$
numerically with high precision. However,one
must be very careful about the order of limits when trying to evaluate
Eq.(\ref{eq:regsum})
numerically in $QCD_{1+1}$ or other theories:
when one employs DLCQ \cite{dlcq} or a similar regulator to
calculate the wavefunctions (and from those the coupling constants
$F_P(n)$) and spectra then it is crucial to send the
DLCQ-cutoff to infinity {\it first} --- otherwise one gets zero
or nonsense for the condensate from the sum rule calculation.
Only {\it after} sending the DLCQ cutoff to zero one may send
the UV-regulator  $\Lambda$ to infinity or the quark mass to zero.

In order to generate an exact result we will use a trick and
replace the sum in Eq.(\ref{eq:regsum}), where both small
and large $n$ contribute, by a sum which is dominated by
large $n$ only. Due to lack of space, only the basic ideas of
the derivation will be sketched here ---
a more detailed discussion of the limit $\Lambda \rightarrow
\infty$ can be found in Ref.\cite{mb:tbp}.
For this purpose, let us consider
\cite{ei:qcd}
\begin{equation}
G(x,\Lambda)\equiv \sqrt{\frac{N_C}{\pi}}
\sum_n \phi_n(x) \left( \frac{1}{M_n^2}-
\frac{1}{M_n^2+\Lambda^2}\right) f_P(n).
\label{eq:defgr}
\end{equation}
Obviously this ``Green's function'' can be used to
compute the condensate, via
\begin{equation}
\left.\langle 0|\bar{\psi}\psi |0\rangle \right|_{ren}
=-m_q^2 \lim_{\Lambda \rightarrow \infty}
\int_0^1 \frac{dx}{x}
\left[G(x,\Lambda)-\left.G(x,\Lambda)\right|_{g=0}\right].
\end{equation}
{}From the equation of motion (\ref{eq:thooft}) one can
show that
\begin{equation}
f_P(n)=\frac{M_n^2}{2m_q}\sqrt{
\frac{N_C}{\pi}}\int_0^1 dy \phi_n(y).
\end{equation}
If one furthermore
invokes completeness of `t Hooft's wavefunctions, i.e.
\begin{equation}
\sum_n \phi_n(x)\phi_n(y) = \delta(x-y)
\end{equation}
one can simplify the first term in $G(x,\Lambda)$,
yielding
\begin{equation}
\sqrt{\frac{N_C}{\pi}}\sum_n \phi_n(x)f_P(n)/M_n^2
=\frac{N_C}{\pi}\frac{1}{2m_q}
\end{equation}
--- independent
of $x$ and the coupling constant. This term thus
drops out completely when we subtract the free field
Green's function.

\begin{figure}
\begin{Large}
\unitlength1.cm
\begin{picture}(14,7.5)(2.3,-10.4)
\put(15.8,-10.3){\makebox(0,0){$m_q/\sqrt{G^2}$}}
\put(3.4,-4.2){\makebox(0,0){$\frac{
\langle 0|\bar{\psi}\psi|0\rangle}{N_C\sqrt{G^2}}$}}
\includegraphics{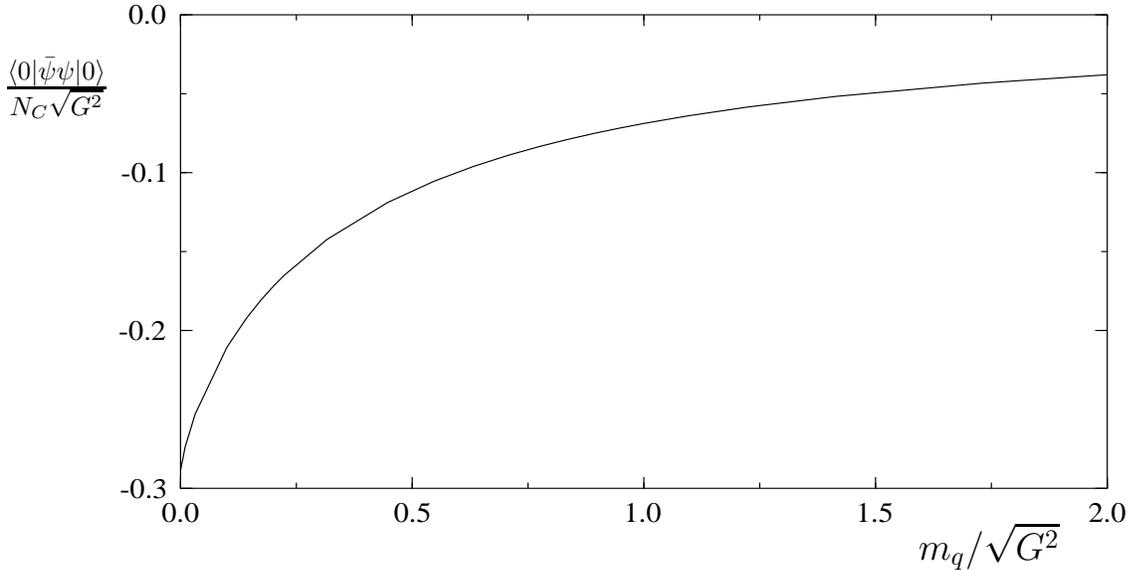}
\end{picture}
\end{Large}
\caption{Renormalized quark condensate as a function of the
quark mass. Both in units of the effective coupling
$G^2=g^2C_F/\pi$. Note the absence of singularities.}
\label{fig:one}
\end{figure}

The crucial point is that, for $\Lambda^2\rightarrow
\infty$, the remaining term in $G(x,\Lambda)$
is dominated by the terms with $n \rightarrow \infty$:
each individual meson yields a negligible
contribution $\propto (M_n^2+\Lambda^2)^{-1}$ when
we send the cutoff to infinity, and a nonzero result
arises only from the summation over infinitely many
highly excited meson states. We have thus succeeded
in converting the {\it low energy sum rule}
(\ref{eq:regsum}) into a
{\it high energy sum rule} and we can now make use
of the abovementioned scaling properties of meson masses $M_n^2$ and coupling
constants $f_P(n)$ as well as of the wavefunction itself
\begin{equation}
\phi_n(x) \stackrel{n \rightarrow \infty}
{\longrightarrow} \Phi(M_n^2x).
\end{equation}
The scaled
wavefunction
\begin{equation}
\Phi(z) \equiv \lim_{n\rightarrow \infty}
\phi_n(z/M_n^2)
\end{equation}
satisfies the integral equation \cite{ei:qcd}
\begin{equation}
\Phi(z) = \frac{m_q^2}{z}\Phi(z) + G^2 \int_0^\infty dy
\frac{\Phi(z)-\Phi(y)}{(x-y)^2}
\label{eq:scaled}.
\end{equation}
In terms of these scaled quantities we thus find
\begin{eqnarray}
&&\left.\langle 0|\bar{\psi}\psi |0\rangle \right|_{ren}
=
\\
&&\frac{N_C}{\pi}m_q^2 \int_0^1 \frac{dx}{x}
\sqrt{G^2}\sum_{n=1,3,..} \frac{\Phi(n\pi^2G^2x)}{n\pi^2G^2+\Lambda^2}
-``g=0''
\nonumber
\end{eqnarray}
Upon performing the substitution $z=n\pi^2G^2 x$,
replacing \footnote{This is exact for $\Lambda \rightarrow \infty$
since the series receives nonvanishing contributions
only from the $n \rightarrow \infty$ region.}
$\sum_{n=1,3,..} \rightarrow (2\pi^2G^2x)^{-1}\int_0^\infty dz$
and performing the $x$-integral one ends up with
\begin{equation}
\left.\langle 0|\bar{\psi}\psi |0\rangle \right|_{ren}
= \frac{N_C}{\pi}m_q^2 \frac{1}{\sqrt{G^2}}
\int_0^\infty \frac{dz \ln z}{z} \Phi(z)
-``g=0''.
\label{eq:logz}
\end{equation}
Note that $\frac{1}{\sqrt{G^2}}
\int_0^\infty \frac{dz}{z} \Phi(z)=\frac{\pi}{m_q}$ \cite{br:scal}
is independent of $G^2$ and it does not matter that the argument
of the logarithm is dimensionful, since a free theory
subtraction is performed.
While `t Hooft's equation cannot be solved
exactly, the scaling equation can be solved
analytically (here we closely follow Ref. \cite{br:scal}):
{}From Eq.(\ref{eq:logz}) it is clear that
$\left.\langle 0|\bar{\psi}\psi |0\rangle \right|_{ren}$
can be related to the Mellin transform
of $\Phi(z)$, for which a closed form expression has been given in
Ref.\cite{br:scal}. The details of the calculation are given in Appendix
\ref{sec:a}. The final result for the renormalized
condensate is (Fig.\ref{fig:one})

\begin{eqnarray}
\left.\langle 0|\bar{\psi}\psi |0\rangle \right|_{ren}
= \left.\frac{m_qN_C}{2\pi}
\right\{ \ln \left(\pi \alpha\right)  -1 -\gamma_E
\quad \quad \nonumber\\
\quad \quad \quad \left. + \left(1-\frac{1}{\alpha}\right)
\left[ (1-\alpha) I(\alpha)-\ln 4\right]\right\},
\label{eq:qexact}
\end{eqnarray}
where $\alpha =G^2/m_q^2$, $\gamma_E =.5772..$ is Euler's
constant and
\begin{equation}
I(\alpha) = \int_0^\infty \frac{dy}{y^2}
\frac{ 1 - \frac{y}{\sinh y \cosh y}}
{\left[\alpha (y \coth y-1)+1\right]}.
\label{eq:exact}
\end{equation}
This result is exact for $N_C\rightarrow \infty$ and
{\it all} quark masses. In the limit $\alpha \rightarrow
\infty$ one recovers Zhitnitsky's result Eq.(\ref{eq:mis0}).
Furthermore, one can verify that
the exact result coincides with the numerical
evaluation of Eq.(\ref{eq:regsum}). In the limit
$\alpha \rightarrow 0$ the condensate vanishes, which
is not surprising since we have subtracted the free
field result.
As one can see from Fig.\ref{fig:one}, $\langle 0|\bar{\psi}\psi |0\rangle$
has an infinite derivative as a function of the quark mass for
$m_q=0$. It arises from a logarithmic singularity in a strong
coupling expansion of Eq.(\ref{eq:qexact})
\begin{equation}
\left.\langle 0|\bar{\psi}\psi |0\rangle \right|_{ren} = -N_C \left[
\sqrt{\frac{G^2}{12}} -\frac{m_q}{2\pi} \ln \left(\frac{m_q^2}{G^2}\right)
+{\cal O}(N_C)\right].
\end{equation}
As will be discussed in the next section, this logarithmic term gives
rise to sizable violations of SU(3) flavor symmetry.
\newpage

This term is interesting for another reason: since the
argument of the logarithm is proportional to the $m_q$
(which itself is proportional to $m_\pi^2$) it looks like
a chiral logarithm. However, chiral logarithms usually arise
from meson loops which are absent here since we work to leading
order in $1/N_C$.

\section{Flavor dependence of the quark condensate}
Even though $QCD_{1+1}$ is not $QCD_{3+1}$ it is interesting
to test some assumptions that are commonly used (but of course
not tested) in the analysis of meson spectra in $3+1$ dimensions
by making (and testing!) the same assumptions in $1+1$ dimensions,
where exact results are available.

In $QCD_{3+1}$ one often assumes that the strange quark condensate
is about the same as the condensate of light quarks even though the
strange quark mass is quite sizable (compared to $\Lambda_{QCD}$).
Based on that (and more) assumption one for example derives the
Gell-Mann--Okubo relation
\begin{equation}
0.988\, (GeV)^2 = 4m_K^2 \approx 3 m_\eta^2 + m_\pi^2 = 0.924\, (GeV)^2,
\end{equation}
which is surprisingly well satisfied, which is often used as a
justification for assuming
\begin{equation}
\langle 0|\bar{s}s| 0\rangle \approx \langle 0|\bar{u}u| 0\rangle.
\label{eq:approx}
\end{equation}
Before one can test the quality of approximations such as Eq.(\ref{eq:approx})
in the case of broken flavor symmetry one must fix the mass scale.
Fixing the slope of the Regge trajectory in the 't Hooft model
yields \cite{roll}
\begin{equation}
G^2=\left( 581 MeV \right)^2.
\end{equation}
The masses of light and strange quarks are fixed by fitting the masses
for $\pi$ and $K$ mesons, yielding
\begin{eqnarray}
m_q \equiv (m_u+m_d)/2 &=& 8.6\, MeV\nonumber\\
m_s &=& 195 MeV \approx \frac{1}{3}G.
\end{eqnarray}
With these quark masses, the Gell-Mann--Okubo mass relation works reasonably
well in $QCD_{1+1}$
\begin{equation}
0.984\, (GeV)^2 = 4m_K^2 \approx 3m_\eta^2+m_\pi^2 = 1.1\, (GeV)^2.
\end{equation}
However, a glance at Fig.\ref{fig:one} shows that the strange
condensate is only about half as large as the light quark condensate
for our values for the quark masses, i.e. the SU(3) flavor symmetry
for the vacuum condensate of the quarks is violated by 50\%.
These results improve somewhat when one uses different criteria to
set the mass scale. For example, a better fit to the meson spectrum
is obtained with $G=1.16 \, GeV$, $m_q=5.8\, MeV$ and
$m_s=130\, MeV$ \cite{roll}. But even for those values there
is still a 25\% violation of the SU(3) symmetry for the vacuum
condensate.

Of course, this is only a toy model and not a serious approximation
to $QCD_{3+1}$ and therefore one should be very careful to draw
any conclusions from this about the real world. Nevertheless,
these results should be taken as a warning and one should be very
cautious about assumptions concerning SU(3) symmetry of quark
condensates.

\section{The gluon condensate}
In $QCD_{1+1}$ gluons are not a dynamical degree of freedom but
appear only in connection with the Coulomb interaction between quarks.
Therefore, it should not be surprising that there is a direct
connection between the quark condensate and the gluon condensate in
$QCD_{1+1}$. In order to find out more, let us first consider the
behavior of the theory under scale transformations
\begin{eqnarray}
\delta \psi &=& \left(\frac{1}{2} + x^\mu\partial_\mu \right) \psi
\nonumber\\
\delta A_\nu &=& \left( 1 +  x^\mu\partial_\mu \right) A_\nu    .
\end{eqnarray}
Since the coupling constant carries dimensions of mass, it
is not surprising that scale invariance is violated ---
even in the
chiral limit
\begin{equation}
\delta \int d^2 {\cal L} = \int d^2 \Delta(x),
\end{equation}
where
\begin{equation}
\Delta(x) = T_\mu^\mu = m \bar{\psi}\psi - \frac{1}{2} F_{\mu \nu}
F^{\mu \nu}.
\end{equation}
The Ward identities for scale transformations imply a
relation between an arbitrary operator $O(x)$ and its
change $\delta O(x)$ under scale transformations
\cite{scale}
\begin{equation}
\langle 0|\delta O |0\rangle = -i\int d^2x \langle 0|
T\left[\Delta(x) O(0)\right]|0\rangle .
\end{equation}
For $O(x)=\bar{\psi}\psi$ this implies
\begin{equation}
i \int d^2x \langle 0| T\left[ \left( m\bar{\psi}\psi(x)- \frac{1}{2}
F^2(x)\right) \bar{\psi}\psi(0)\right]|0\rangle =
- \langle 0|\bar{\psi}\psi |0\rangle.
\end{equation}
When one combines this result with
\begin{equation}
\frac{d}{dm} \langle 0|F^2|0\rangle
= -i \int d^2x \langle 0 |T\left[ \bar{\psi}\psi(x) F^2(0)\right]|0\rangle
\end{equation}
and
\begin{equation}
\frac{d}{dm} \langle 0|\bar{\psi}\psi|0\rangle
= -i \int d^2x \langle 0 |T\left[ \bar{\psi}\psi(x)
\bar{\psi}\psi(0)\right]|0\rangle
\end{equation}
one obtains
\begin{equation}
\frac{d}{dm} \langle 0|F^2|0\rangle =
2m^2 \frac{d}{dm} \frac{\langle 0|\bar{\psi}\psi|0\rangle}{m},
\label{eq:ddmf}
\end{equation}
which can be easily used to express the gluon condensate in terms
of the quark condensate --- up to a quark mass independent constant.
Notice that the quark mass dependent piece of the gluon condensate
is in fact only of $\cal{O}(N_C)$, which should, however,
not come as a surprise since the gluons appear only through the
quarks in $QCD_{1+1}$.
Using the exact result for the quark condensate, it is a straightforward
calculation to generate an exact expression for the quark mass dependent part
of
the gluon condensate in $QCD_{1+1}$ from Eq.(\ref{eq:ddmf}) .
However, since it was not possible to determine the integration
constant, the result will not be displayed here.

Besides the quark and gluon condensate, one can also consider
mixed condensates in $QCD_{1+1}$ \cite{chib}. Of particular interest are
combinations like $\langle 0|\bar{q} (x^\mu D_\mu)^nq|0\rangle$ because
of the connection to the propagator of a heavy-light system \cite{chib}.
However, I was not able to derive exact results for such combinations.

\section{Discussion}
The exact result which we have obtained (\ref{eq:exact})
is an analytic function of $\alpha$ in the complex
plane cut along the negative real axis --- i.e. there
is {\it no} phase transition. An asymptotic expansion
in powers of $\alpha$ yields
\begin{equation}
\frac{2\pi}{m_q N_C} \left.\langle 0|\bar{\psi}\psi |0\rangle \right|_{ren} =
\sum_{\nu=1}^\infty c_\nu \alpha^\nu,
\label{eq:asym}
\end{equation}
where the coefficients show factorial growth
\begin{equation}
c_\nu \stackrel{\nu \rightarrow \infty}{\sim}
(-1)^\nu e^{-2} 2^{1-\nu}(\nu-1)!\quad,
\end{equation}
i.e., the asymptotic expansion for
$\left.\langle 0|\bar{\psi}\psi |0\rangle \right|_{ren}$
is only Borel summable and the Borel series has a
finite radius of convergence.
Applying the inverse Borel transfer to the Borel
summed series one recovers the exact result which
reflects the absence of terms like $e^{-\frac{1}{\alpha}}$.
I have compared the first three terms in the
asymptotic expansion with the perturbative (Feynman diagrams) expansion and
found agreement.
Nevertheless Eq.(\ref{eq:exact}) is a completely
nonperturbative result,
because one has to sum up {\it all} terms in the
perturbation series before one obtains the right
scaling behavior (\ref{eq:mis0}) for small quark mass
(large $\alpha$). It is also {\it not} sufficient to
keep only the asymptotic behavior of the series: it is easy to write down an
expression which has the same asymptotic
coefficients for large $\nu$ but does not yield the desired
$\sqrt{\alpha}$ behavior for $m_q\rightarrow 0$:
\begin{equation}
f(\alpha)= - 2\alpha e^{-2} \int_0^\infty dy \frac{e^{-y}}{2+\alpha y}
\end{equation}
has the same large $\nu$ behavior for the asymptotic series as the r.h.s. of
Eq.(\ref{eq:asym}).
However, in the strong coupling limit,
$f(\alpha) \sim \alpha \ln \alpha$, i.e. looking only at the
tail of the asymtotic series yields a strong coupling behavior which is too
singular. There is also another way to see that
the asymptotic behavior of the series has little do do with the behavior
of the exact result in the strong coupling limit:
Both in Eq.({\ref{eq:exact}) as well as in $f(\alpha)$, the $y \rightarrow 0$
region of the integral are crucial for the behavior in the
strong coupling limit but it is the $y \rightarrow \infty$ part of the
integral that is responsible for the behavior of the asymptotic series.

We have started from `t Hooft's equation which is based
on light-front quantization. The light-front vacuum
is trivial, i.e. identical to the Fock space vacuum \cite{le:ap}.
Nevertheless, using current algebra and sum rule
techniques, we obtained a nonzero result for the
quark condensates. The result we obtained agrees with
numerical calculations using equal time quantization
(see Refs.\cite{le:ap,li:vac} for $m_q=0$ and
Ref.\cite{mth:priv} for the general case).
The apparent paradox (nontrivial condensates from
trivial vacua) is clarified by defining the LF field theory
with its light-like quantization hypersurface
through a limiting procedure in which the
quantization hypersurfaces are spacelike \cite{le:ap,horn}.
The basic result from such studies \cite{mb:sg,naus}
is that the vacuum in LF front field theory appears to
be frozen and correct spectra and structure function
(leading twist) results are obtained by solving
a suitable LF Hamiltonian which has a trivial vacuum.
For some quantities, however, the limiting transition to
the LF is not smooth and by working directly on the LF
one obtains incorrect results.
The quark condensate is such an example.
On the other hand, the sum rule calculation based on
(leading twist) LF wavefunctions yields the correct results
since spectrum and (leading twist) wavefunctions show
a smooth LF limit in $QCD_{1+1}$.

\appendix
\section{Exact expression for the Mellin transform of the scaling function}
\label{sec:a}
For $-\beta < \lambda <1$ one can introduce the Mellin transformed function
\begin{equation}
\Psi(\lambda) = \int_0^\infty dz z^{\lambda-1} \Phi(z).
\end{equation}
{}From Eq.(\ref{eq:logz}) it is clear that , in order to calculate the
condensate,
it is sufficient to calculate $\Psi^\prime(0)$, using
\begin{equation}
\left.\langle 0|\bar{\psi}\psi |0\rangle \right|_{ren}
= \frac{N_C}{\pi}m_q^2 \left[ \frac{1}{\sqrt{G^2}}
\Psi^\prime (0)
-``g=0'' \right].
\end{equation}
The Mellin transform $\Psi(\lambda)$ satisfies the difference equation
\begin{equation}
\Psi(\lambda+1) = \left( \pi \lambda \cot \pi \lambda - \pi \beta \cot \pi
\beta
\right) \Psi(\lambda).
\label{eq:diff}
\end{equation}
The analytic continuation of $\Psi(\lambda)$ yields a meromorphic function
with poles at $\lambda=2,3,4,...$ as well as $\lambda=-\beta_0,-1-\beta_1,
-2-\beta_2,...$, where $\beta_n \in (0;1)$ is the unique solution of
\begin{equation}
\pi(n+\beta_n) \cot \pi \beta_n = 1-\frac{1}{\alpha}.
\end{equation}
The solution to this difference equation (\ref{eq:diff}) is given in
Ref.\cite{br:scal}
\begin{equation}
\Psi(\lambda) =\frac{1}{\sqrt{\alpha}}\Psi_0(\lambda) \prod_{n=0}^\infty
\frac{1+ \frac{m_q^2-G^2}{G^2 \pi(\beta_n+n)} \tan \pi \lambda}
{1+ \frac{m_q^2-G^2}{G^2 \pi(\lambda+n)} \tan \pi \lambda},
\end{equation}
where
\begin{equation}
\psi_0(\lambda) = \pi^\lambda \Gamma(\lambda) \exp \left[
-2\pi \int_0^{\lambda-1} du \frac{u+\frac{1}{2}\sin^2\pi u}{\sin (2\pi
u)}\right]
\end{equation}
is the solution of the difference equation for $m_q^2=G^2$ (i.e. $\alpha=1$).
The overall normalization (which is not determined from the difference
equation alone since Eq.(\ref{eq:diff}) is linear in $\Psi$).
It has been fixed using the known scaling behavior of the pseudoscalar
coupling constants $\lim_{n\rightarrow \infty}m_q\int_0^1 dx
\frac{\phi_n(x)}{x}
 =\pi G$, which implies
\begin{equation}
\Psi(0) \equiv \int_0^\infty dz \frac{\Phi(z)}{z} = \pi \sqrt{\alpha}.
\end{equation}
{}From these results it is now straightforward to evaluate the derivative of
$\Psi(\lambda)$ at the origin, yielding
\begin{eqnarray}
\Psi^\prime(0) =& \left.\frac{\pi}{m_q}\right\{\ln \pi -1 - \gamma_E\quad \quad
\quad \quad \quad \quad \quad \quad \quad
\label{eq:sumrep}
\\
&\quad \quad \left.+ (1-\frac{1}{\alpha})\left[\frac{1}{\beta_0}
+\sum_{n=1}^\infty \left(\frac{1}{n+\beta_n}-\frac{1}{n}\right)\right]\right\}
{}.
\nonumber
\end{eqnarray}
This expression is most convenient for $m_q^2 \ll G^2$ (i.e.
$\alpha \rightarrow \infty$) since there $\beta_0\rightarrow 0$ and
the r.h.s. of Eq.(\ref{eq:sumrep}) is dominated by $1/\beta_0$.
For the general case it is more convenient to use an integral representation
(which can easily be derived from Eq.(\ref{eq:sumrep}) using contour
integration)
\begin{eqnarray}
\Psi^\prime(0) = &\left. \frac{\pi}{m_q}\right\{\ln \pi -1 -\gamma_E
\quad \quad \quad \quad \quad \quad \quad \quad
\\
&\quad \quad \left.+ (1-\frac{1}{\alpha})\left[-\ln 4 + (1-\alpha)I(\alpha)
\right]
\right\},
\nonumber
\end{eqnarray}
where $I(\alpha)$ has been defined in Eq.(\ref{eq:exact}).


\begin{references}
\bibitem{zhit:vac} A. R. Zhitnitsky, {\it Phys. Lett.} {\bf 165B}, 405 (1985);
{\it Sov. J. Nucl. Phys.} {\bf 43}, 999 (1986); {\it ibid.}
{\bf 44}, 139 (1986).
\bibitem{co:no} S. Coleman, {\it Comm. Math. Phys.} {\bf 31}, 259
(1973).
\bibitem{thooft} G. 't Hooft, {\it Nucl. Phys.} {\bf B75}, 461 (1974).
\bibitem{mb:tbp} M. Burkardt, talk presented at ``Quantum Infrared Physics'',
eds. H.M. Fried and B. M\"uller, Paris, June 1994, hep-ph/9409333.
\bibitem{ei:qcd} M. B. Einhorn, {\it Phys. Rev.} {\bf D14}, 3451
(1976).
\bibitem{dlcq} T. Eller, H.-C. Pauli and S.J. Brodsky, {\it Phys. Rev.} {\bf
D35}, 1493 (1987).
\bibitem{br:scal} R. C. Brower, W. L. Spence and J. H. Weis,
{\it Phys. Rev.} {\bf D19}, 3024 (1979).
\bibitem{roll} W.B.Rollnik, {\it Phys. Rev.} {\bf D17}, 3086 (1978).
\bibitem{scale} S. Coleman in 1971 Erice summer school, (Editrice Compositori,
Bologna, 1973);
V.A. Novikov, M.A. Shifman, A.I. Vainstein and V.I. Zakharov,
{\it Nucl. Phys.} {\bf B191}, 301 (1981).
\bibitem{chib} B. Chibisov and A. R. Zhitnitsky, hep-ph/9502258.
\bibitem{le:ap} F. Lenz, S. Levit, M. Thies and K. Yazaki,
{\it Ann. Phys. (N.Y.)} {\bf 208},1 (1991).
\bibitem{mb:sg} M. Burkardt, {\it Phys. Rev.} {\bf D47}, 4628 (1993);
M. Burkardt, {\it to appear in Adv. Nucl. Phys.} {\bf 23}, hep-ph/9505259.
\bibitem{naus} E. Prokhvatilov, H.W.L. Naus and H.J. Pirner,
{\it Phys. Rev.} {\bf D51}, 2933 (1995); J.P. Vary and T.J. Fields,
talk given at Workshop on ``Theory of Hadrons and Light-Front QCD'', Polona
Zgorzelisko, Poland, 15-25 Aug 1994, ed. S. Glazek, hep-ph/9411263.
\bibitem{li:vac} M. Li, {\it Phys. Rev.} {\bf D34},3888 (1986).
\bibitem{mth:priv} M. Thies, {\it private communications}.
\bibitem{horn} K. Hornbostel,  {\it Phys. Rev.} {\bf D45}, 3781 (1992).
\end{references}
\end{document}